\documentclass{lps} 
\usepackage{lipsum}                   
\usepackage[round]{natbib} 
\usepackage{url}

\begin{document}
\title{Are Programming Paradigms Paradigms?}
\Subtitle{A Critical Examination of Floyd's Appropriation of Kuhn's Philosophy}
\author{Peyman M. Kiasari}
\shortauthor{Peyman M. Kiasari}

\maketitle
\tableofcontents


\begin{abstract}
This paper examines the philosophical relationship between Thomas Kuhn's concept of scientific paradigms and the programming paradigm concept in computing that was introduced by Floyd in his 1978 Turing Award lecture. Through critical analysis of both Kuhn's original framework and its application in computing, we argue that the contemporary usage of `programming paradigms' represents a significant departure from Kuhn's philosophical concept. We demonstrate that while Floyd explicitly attributed this term to Kuhn's work, his usage fundamentally altered the concept's meaning. We argue that this divergence necessitates a critical reassessment of the term's usage in computing discourse.
\end{abstract}

\section{Introduction}

The term `paradigm' has become deeply embedded in computing discourse since its introduction by Robert Floyd in his 1978 Turing Award lecture \citep{Floyd1979}. This influence is evidenced by its widespread adoption in foundational computing literature \citep{SICP, PLPP, CTM} and its persistent use in describing programming methodologies. In his lecture, Floyd sought to address what he perceived as inadequacies in programming methodology and education, arguing that focusing solely on language syntax and algorithms was insufficient. He proposed developing and teaching explicit programming `paradigms' as a more comprehensive approach to software development. Floyd explicitly attributed his inspiration to Thomas Kuhn's \textit{The Structure of Scientific Revolutions} \cite{Kuhn1962}. However, his understanding of the concept contrasted substantially with Kuhn's original framework, initiating a separate trajectory of the term in computing.

Our analysis proceeds in four stages. First, we examine Kuhn's original conception of paradigms and its key characteristics. Second, we critically evaluate Kuhn's insistence on paradigm singularity. Third, we analyze Floyd's adaptation of the concept in computing. Finally, we assess the philosophical implications of this conceptual translation.

\section{Related Work}

The philosophical examination of programming paradigms has centered largely around two key questions: first, whether programming paradigms genuinely are paradigms in Kuhn's sense, and second, how the concept of paradigm has been transformed in its application to computing. Several scholars have contributed important perspectives to this discourse.

Harper \citep{Harper2017} provides an alternative perspective on programming paradigms by drawing an analogy to Gould's critique of zebra taxonomy. Just as Gould argued that genomics provides a more fundamental understanding than morphological classification of zebras, Harper suggests that type theory, rather than paradigmatic divisions, provides the foundational ``genomics" of programming languages.

Michaelson \citep{Michaelson2020} demonstrates that the apparently different paradigms like procedural, object-oriented, and functional programming exhibit substantial theoretical overlap and methodological continuity. Rather than viewing these as distinct paradigms, he argues that the coexistence of multiple paradigms contradicts Kuhn's emphasis on paradigm singularity and they are better understood as different traditions within a unified ``computer science paradigm". However, Michaelson's analysis does not fully address the fundamental philosophical tensions between these traditions: the functional paradigm's rejection of mutable state versus the imperative paradigm's embrace of it, and certain type systems'  rejection of null values, and other tensions that suggest deeper incompatibilities-it's anarchistic rather than paradigmatic.

A contrasting perspective emerges from van Roy's \citep{vanRoy2009} work, which identifies thirty distinct programming paradigms based on formal operational properties. This taxonomic approach is further developed in his comprehensive treatment with Haridi \citep{CTM}. However, as Krishnamurthi \citep{Krishnamurthi2008} argues, such taxonomic approaches, while valuable for understanding language features, may misconstrue the nature of paradigmatic distinctions. His work suggests that the diversity of programming approaches represents ``as different traditions, each offering its own perspective on computational ideas".

\section{Kuhn's Paradigm Concept}

This section examines Kuhn's concept of scientific paradigms and its implications for understanding scientific development. We begin by analyzing Kuhn's characterization of paradigms and their role in scientific progress, followed by a critical evaluation of his assertion regarding paradigm singularity. This examination reveals important considerations for applying Kuhn's framework to the term ``Programming Paradigms".

\subsection{Examination}

A paradigm can be understood as a framework of ideas, theories, and methods that shapes how a field understands and investigates its subject matter. Kuhn provides a more specific characterization in his work, stating that ``close historical investigation of a given specialty at a given time discloses a set of recurrent and quasi-standard illustrations of various theories in their conceptual, observational, and instrumental applications. These are the community's paradigms" \citep[p. 43]{Kuhn1962}.

In Kuhn's analysis of scientific development, the progression of science follows a characteristic pattern: pre-paradigm  transforms into normal science, which may be disrupted by extraordinary science (model crisis and paradigm shift), potentially leading to a new normal science phase. During pre-paradigm science, multiple competing schools of thought coexist, each with its own theoretical foundations and methodological approaches. The transition to normal science occurs when a single paradigm achieves widespread acceptance within the scientific community.

The dominance of a single paradigm during periods of normal science represents a fundamental aspect of Kuhn's framework. This singularity is so essential that Kuhn considers it a defining characteristic of science itself. As Bird notes in his analysis, ``To achieve the status of a science, a discipline must reach consensus with respect to a single paradigm" \citep{IEPKuhn}. This emphasis on paradigmatic unity shapes Kuhn's entire conception of scientific development.

Kuhn's insistence on paradigm singularity is further evidenced by his treatment of paradigmatic diversity. While he acknowledges that ``there are circumstances, though I think them rare, under which two paradigms can coexist peacefully" \citep[p. ix]{Kuhn1962}, this concession is presented as an exceptional case that proves the rule. More tellingly, Kuhn explicitly associates the presence of fundamental debates and methodological disagreements with pre-paradigm periods rather than mature science: ``The pre-paradigm period, in particular, is regularly marked by frequent and deep debates over legitimate methods, problems, and standards of solution, though these serve rather to define schools than to produce agreement" \citep[pp. 47-48]{Kuhn1962}. This characterization reinforces his view that genuine scientific practice requires convergence.

\subsection{Critique of Kuhn's Single Paradigm Thesis}

Whatever one might disagree on, we can't disagree that disagreements exist in science. This reality is evident across scientific disciplines and throughout history, for example the ongoing debate ``Scaling vs. Inductive Bias" that has divided deep learning researchers into distinct methodological camps. Given this observable phenomenon of scientific disagreement, Kuhn's insistence on paradigm singularity requires philosophical justification.

Kuhn does acknowledge the existence of competing viewpoints in science, stating that ``There are schools in the sciences which approach the same subject from incompatible viewpoints..." \citep[p. 177]{Kuhn1962}. However, his response to this challenge is philosophically problematic: ``...But they are far rarer there than in other fields; they are always in competition; and their competition is usually quickly ended" \citep[p. 177]{Kuhn1962}. This response bears characteristics of the No True Scotsman fallacy—when presented with counterexamples, Kuhn simply asserts they must be ``rare'' or ``quickly ended'' without providing criteria for what constitutes ``rare'' or ``quick.'' The argument becomes circular: genuine paradigms must be singular because situations with multiple approaches are defined as not being true paradigm-governed science.

This potential philosophical weakness in Kuhn's framework adds complexity to our central inquiry into whether programming paradigms are paradigms. If we cannot fully accept his insistence on paradigm singularity we must evaluate programming methodologies against a framework whose philosophical foundations we have found to be problematic.

\section{Floyd's Programming Paradigms}

This section analyzes Floyd's interpretation and application of the paradigm concept in programming, followed by a critical evaluation of his approach. The analysis highlights key distinctions between Floyd's use of paradigmatic thinking and Kuhn's original philosophical framework.

\subsection{Examination}

Floyd's central thesis emerges from his concern about the state of programming education and practice. He argues that the contemporary focus on teaching programming languages' syntax and algorithms is insufficient, proposing instead that programming education should center on teaching explicit programming ``paradigms". In making this argument, Floyd explicitly draws on Kuhn's work, stating: ``Thomas S. Kuhn, in The Structure of Scientific Revolutions, has described the scientific revolutions of the past several centuries as arising from changes in the dominant paradigms. Some of Kuhn's observations seem appropriate to our field".

One of his key examples is structured programming, which he describes as ``the dominant paradigm in most current treatments of programming methodology". He explains that structured programming consists of two phases: top-down design (where problems are decomposed into simpler subproblems) and working upward from concrete objects to more abstract functions. This example demonstrates how Floyd conceptualizes paradigms as practical methodological approaches to program design.

The lecture concludes with three distinct calls to action: he advises programmers to examine and refine their methods, encourages teachers to explicitly identify and teach their paradigms, and urges language designers to support the paradigms used by programmers. Floyd argues that the advancement of programming requires the ``continuing invention, elaboration, and communication of new paradigms," positioning paradigms as essential tools for addressing what he terms the ``software depression" - a persistent state of unreliability, inefficiency, and unresponsiveness in software development.

\subsection{Critique of Floyd's Programming Paradigms Thesis}

Floyd's lecture begins, notably, with a dictionary definition of the word ``paradigm" as ``pattern, exemplar, example." This starting point is noteworthy not only for what it includes but what it omits—while Floyd would later explicitly reference Kuhn's work, his initial framing through the dictionary definition suggests a more general, less philosophically specific interpretation of the term than Kuhn's complex theoretical framework.

Furthermore, the very title of Floyd's lecture—``The Paradigms of Programming"—with its use of ``paradigms" in the plural form, appears to stand in tension with Kuhn's emphasis on paradigm singularity during periods of normal science. While Kuhn does indeed employ the plural ``paradigms" in his work, he does so to describe successive paradigms that emerge through historical progression—paradigms that exist in series, with each new paradigm replacing its predecessor during scientific revolutions or paradigm-shifts. Floyd, in contrast, presents multiple paradigms as not only coexisting but fundamentally complementary, suggesting a parallel rather than serial relationship between different programming approaches. This parallel view of paradigms diverges from Kuhn's framework, where multiple paradigms typically signal either a pre-paradigmatic state or a period of crisis and revolution, rather than a stable, mature field of practice.

To better understand Floyd's concept of `paradigm,' his distinct uses of the term warrant careful examination. He refers to structured programming as ``a familiar example of a paradigm of programming," describing it as ``the dominant paradigm in most current treatments of programming methodology." He identifies ``simultaneous assignment of new values to the components of state vectors" as another paradigm. Perhaps most strikingly, he characterizes merge sorting as ``an instance of the divide-and-conquer paradigm," and refers to the ``state-machine paradigm" as a general approach to computation.

These applications of the term reveal a fundamental divergence from Kuhn's conception. When Floyd describes structured programming as a paradigm, he is referring to a specific methodology for program design and implementation. While this methodology certainly influences how programmers approach their work, it operates at a far more specific level than Kuhn's paradigms, which encompass entire worldviews and ways of understanding reality within a scientific discipline. More strikingly, characterizing merge sorting as an instance of a paradigm reduces the concept to the level of specific problem-solving techniques—a far cry from Kuhn's comprehensive frameworks that shape entire scientific disciplines. Even more telling is Floyd's description of the ``state-machine paradigm," which he presents as one of many possible approaches to computation; Floyd's using the term in a way that more closely aligns with his opening dictionary definition than with Kuhn's philosophical concept.

This divergence from Kuhn's framework becomes particularly apparent in Floyd's treatment of paradigms as teaching tools. While Kuhn saw paradigms as emerging naturally from scientific practice, Floyd presents them as explicit methodologies to be taught and consciously adopted. This prescriptive approach to paradigms, for one who believes computing is science, is paradoxical.

\section{Philosophical Implications}

This examination reveals two significant philosophical implications. First, programming paradigms do not align with Kuhn's framework of scientific paradigms. In modern computing, approaches like object-oriented, functional, and procedural programming coexist and sometimes complement each other, rather than one becoming dominant as Kuhn's paradigms would. If we were to apply Kuhn's criteria strictly, we at least would have to conclude that computing remains in a ``pre-paradigm" state, thus is not mature science.

Second, Floyd's usage of the term `paradigm,' while explicitly drawing from Kuhn, represents a fundamental misappropriation of the philosophical concept. This misuse has propagated and deeply embedded throughout computing literature, where it remains remarkably vague and loosely defined. In both Floyd's work and subsequent computing literature, the term has become so elastic that it can encompass various concepts like patterns, methodologies, principles, approaches, models, and algorithms—serving as a placeholder term without concrete meaning, indicating existence of a common pattern.
\section{Conclusion}

This paper has demonstrated that the term `programming paradigms,' despite its widespread adoption in computing, represents a significant departure from Kuhn's original philosophical framework. While Floyd explicitly drew from Kuhn's work, his adaptation fundamentally altered the concept's meaning. This misalignment suggests that the computing field may benefit from more clear terminology, such as `methodologies', `approaches,' or `principles,' that better reflects how these concepts actually function in practice.

In practical reality, computing tends to be more free-form and practical, mixing different approaches based on what works best for each situation, where programmers freely combine different methods and break traditional rules when needed: this is more anarchistic rather than paradigmatic.

\bibliographystyle{plainnat}  
\bibliography{references}

\end{document}